\documentclass[conference]{IEEEtran}

\IEEEoverridecommandlockouts

\usepackage[numbers]{natbib}

\usepackage{graphicx}

\begin{document}
%
\title{Bluetooth based Proximity, Multi-hop Analysis and Bi-directional Trust: Epidemics and More \thanks{This paper is based on our work led by Ramesh Raskar and Sai Sri Sathya at Facebook in 2016 \newline
US10149136B1 - Proximity-based trust \newline
US10685078B2 - Content provision based on geographic proximity \newline
US20190215753A1 - Proximity-based messaging protocol \newline
EP3512232A1 - Method, computer readable storage media and apparatus for proximity-based trust \newline
US20190207819A1 - Determining Mesh Networks Based on Determined Contexts
}}

\author{\IEEEauthorblockN{Ramesh Raskar}
\IEEEauthorblockA{MIT Media Lab \\ 
Email: raskar@mit.edu}
\and
\IEEEauthorblockN{Sai Sri Sathya}
\IEEEauthorblockA{S20.AI \\
Email: origin@s20.ai}}


%


\maketitle

\begin{abstract}
In this paper, we propose a trust layer on top of Bluetooth and similar wireless communication technologies that can form mesh networks. This layer as a protocol enables computing trust scores based on proximity and bi-directional transfer of messages in multiple hops across a network of mobile devices. We describe factors and an approach for determining these trust scores and highlight its applications during epidemics such as COVID-19 through improved contact-tracing, better privacy and verification for sensitive data sharing in the numerous Bluetooth and GPS based mobile applications that are being developed to track the spread. 
\end{abstract}


%
\IEEEpeerreviewmaketitle

\section{Introduction}
The modern day pandemic, COVID-19 \cite{Source1} has turned the lives of people all over the globe upside down and has pushed experts to panic-hunt for quick-fix solutions in practically every field: healthcare services, epidemiology, social services, economics or policy. It is no wonder, in fact practical, that several governments, institutions and even corporations are coming up with solutions based on GPS and Bluetooth for contact tracing \cite{Source2} \cite{Source3} as the pandemic reaches the community phase of transmission in most countries. The latest of such being the announcement of Google and Apple joining hands for developing Bluetooth-based contact tracing technology \cite{Source4} \cite{Source5} \cite{Source6}. 

While the urgency of a solution is well-warranted, we do understand that Bluetooth and related proximity technologies alone are not enough to solve the problems of contact tracing and that they require additional context \cite{Source7} along with a widespread adoption to be effective which could be very challenging in many countries. Hence, a conscious implementation of the same would go a long way in maintaining the effectiveness of the solution and prevent any negative repercussions that may arise once the solutions are adopted and implemented on-ground. In this paper, we propose the development of a trust layer as a protocol on top of proximity technologies \cite{Source8} like Bluetooth that is bi-directional, recursive and mimics the human-like trusts scenarios between devices in a way that adjusts trust scores based on their previous and current interactions and can transit in multiple hops \cite{Source9} \cite{Source10}. This can increase the effectiveness of contact tracing when there is mass adoption, enhance privacy, and enable contextual message-passing based on proximity information \cite{Source11} during pandemics. 

\section{Background}
Interpersonal human interactions tend to be subjective and the level of trust we place in someone is revised as we continue to have more and more interactions with them. Studies \cite{Source12} \cite{Source13} \cite{Source14} \cite{Source15} have shown that proximity interactions have profound impact on trust formation and development among peers. While there are different types of proximities such as cognitive, social, institutional, cultural, etc., that have an impact on trust, geographical proximity is often considered as the nodal point that coincides and impacts the other types \cite{Source16}. Further studies have been carried out and frameworks have been proposed to model human mobility patterns \cite{Source17} \cite{Source18} \cite{Source19} \cite{Source20} \cite{Source21} and determine context \cite{Source22} \cite{Source23} for building people-centric services \cite{Source23} \cite{Source24} using multi-hop proximity based technologies \cite{Source25}.

We build on these models to propose a universal trust protocol that is based on proximity and can be used for facilitating digital interactions in the physical world that involve trust such as sharing sensitive information, payments, etc. 

Our approach is novel in demonstrating the physics of digital relations based on trust evolving in the real world. We propose that all digitally connected systems embed this trust protocol layer to determine trust scores that adjust over-time and enable information transfer through a multi-hop peer-to-peer wireless network based on trust.

\section{Overview}
Our system is based on proximity over space and time to establish digital trust between users using their mobile devices connected via wireless mesh networks. When a mobile device identifies a trigger for an offline digital communication with one or more devices, it discovers the other devices in proximity using a discovery protocol. For each discovered peer, the device assesses whether the discovered peer is previously known and whether it is directly reachable over the wireless network. It then determines a trust score for each discovered peer based on proximity and related factors and performs digital communications with one or more peers with trust scores higher than a threshold for transmission. The system is bi-directional in nature. So correspondingly, each receiver also computes a trust score for the sender and can decode the message only if the trust score is higher than the threshold for reception. Figure 1 illustrates message passing between two devices, as detailed in section 4.6.

\begin{figure}[htp]
    \centering
    \includegraphics[scale=0.2]{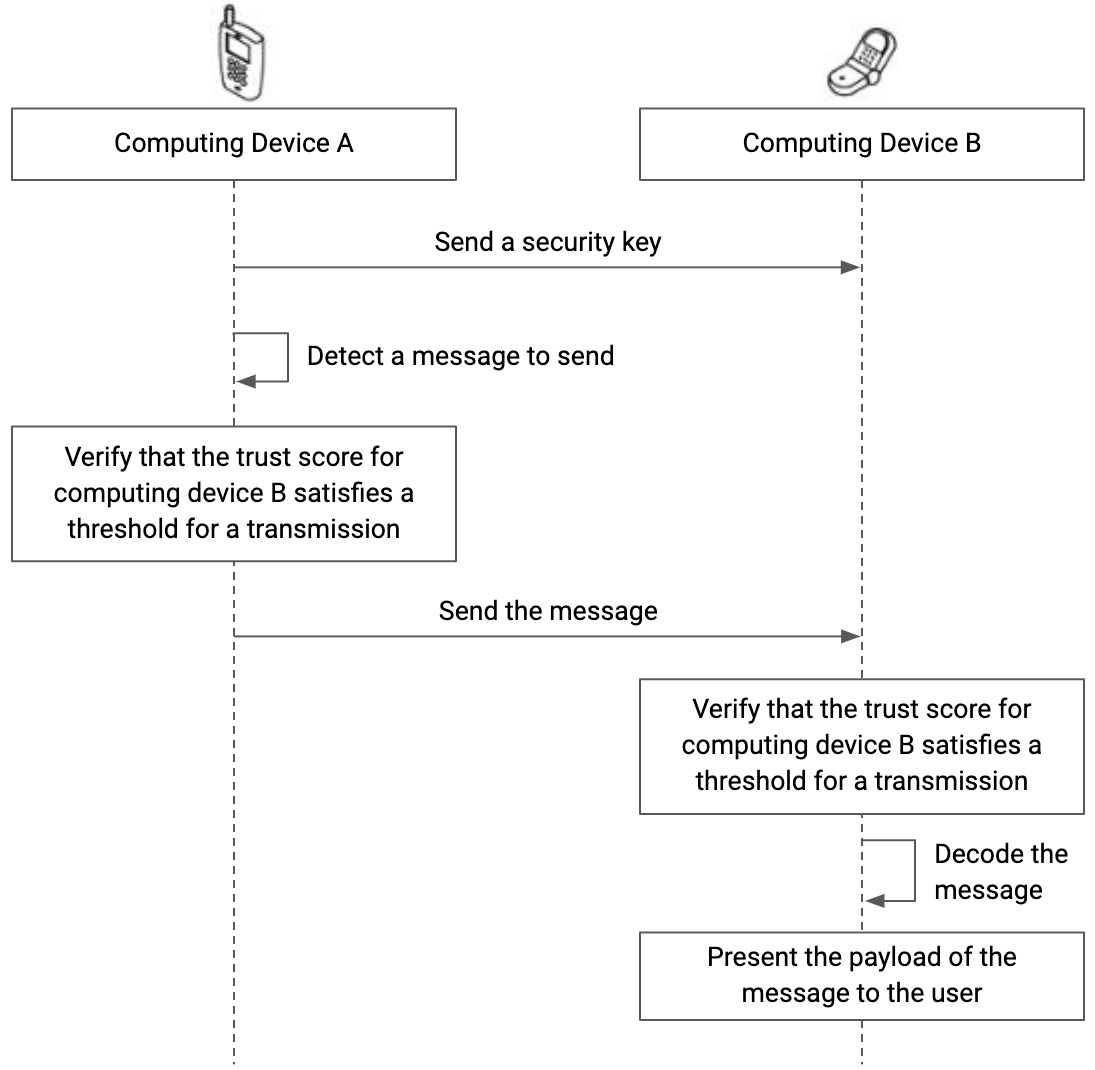}
    \caption{Message passing between two devices based on computed trust scores}
    \label{fig:graphic1}
\end{figure}

\section{Approach}
The following subsections explain the architecture of the system to enable communication between two devices. A triggering event initiates a proximity-based communication between two or more devices. The data captured during the initial interaction is analysed by the devices in various ways to compute initial trust scores. The trust scores are adjusted based on the communication and physical interactions captured and form the basis for any information exchange that takes place between the two devices. 

The communicating devices have one or more sensors that can capture physical interactions of the users. These devices may also communicate with multiple communication peers through different wireless networks simultaneously.

\subsection{Triggers for Offline Communication}
The triggering event for a device to initiate an offline communication with other devices may simply be an instruction from the user or encoded as a set of events. For instance, the user of a mobile device may detect that no network infrastructure is reachable or that multiple peers are in physical proximity to one another. This detection becomes a trigger for initiating an offline peer-to-peer communication session. For example, when a user is on a plane, the mobile device(s) she is carrying should be in the airplane mode. Thus, the mobile devices on the plane are not capable of accessing the Internet. 

The devices only need to exchange digital data among themselves. In this scenario, the devices form a wireless mesh network and exchange messages between themselves without routing the messages through the Internet. 

\subsection{Discovery}
Once an offline digital communication is triggered, the user’s device discovers the other mobile device(s) in proximity. Any existing discovery protocol such as the Bluetooth Low Energy (BLE) discovery protocol can be re-used for this purpose. It can further identify the mobile devices that are interested in communicating with it on a particular topic to initiate a peer-to-peer communication session based on the information in their advertising packets or retrieved using an online centralized or decentralized social network that the one or more peers are a part of including the primary user initiating the communication. For example, a user enters a coffee shop and wants to chat about coffee while he is waiting for his friends. Either the people in the coffee shop could broadcast this information or he could connect to a social network and retrieve a list of devices mapped to users in the coffee shop who would be interested to chat about coffee. This would prompt the user to send local invitations to the users identified through offline and online discovery channels.

\subsection{Types of Communication Sessions}

\begin{itemize}
    \item \textbf{Offline communication session}: where a device exchanges messages with one or more mobile devices without routing through the Internet.  
    \item \textbf{Online communication sessions in proximity}: where a mobile device exchanges messages with one or more mobile devices by routing the messages through the Internet. Though messages are routed through the Internet, the participating mobile devices and their corresponding users could be in proximity of each other. Therefore, in such cases, proximity-based trust is utilized for authorizing the communications. 
    \item \textbf{Hybrid communication session}: where one or more participating mobile devices are not in the same local network or directly connected with each other via the internet. To establish communication, mobile devices having local communication paths could route messages through one or more peers connected to the Internet.
\end{itemize}

As devices use low power radio for the offline communications, not all the devices may be directly accessible to each other even if they are in close proximity. In such cases messages are passed in multiple hops through nearby devices. For example, for two devices A and B; when device A is not reachable from device B, device B sends a message to device A through a third device that is reachable from device B and is able to reach device A. When only a portion of participants is capable of accessing the Internet, they act as backhaul points and route messages from / to the other participants to / from nodes outside the mesh network. Device A may communicate with a second mobile device (device B) over the Bluetooth network while communicating with a third mobile device over the Wi-Fi network. A back-haul point can be one of the participating mobile devices. It can also be a stationary infrastructure device including a Wi-Fi access point. 

Additionally, devices could be associated with one or more centralized or decentralized social-networking system(s). In each of the above cases, the mobile devices could utilize data available in the data stores of the social network when the mobile devices discover each other and maintain the communication session as they move.

\subsection{Analysing Proximity Data and Physical Interactions}
Proximity information and physical interactions are captured through wireless transceivers and on device sensors such as microphone, a camera, etc. Physical interactions between users include conversations, handshakes, hands waving, and any other human interactions that can be captured by any available sensors. To process the data and compute trust scores, on-device machine learning or deep learning (ML/DL) models are used. If the device is connected to the internet, cloud-based servers can be additionally used to process the data. \cite{Source26}

\subsection{Factors for Determination Trust Scores}
Once a device (A) has identified a communication peer (device B) following factors are used for determining trust scores as shown in figure 2:

\begin{itemize}
    \item \textbf{Previous sessions}: If device A and B have interacted within a specified time-frame, then device A would use the previously computed trust score for device B as an initial trust score.
    \item \textbf{Mutual peers}: Trust scores are uniquely computed and stored between each pair of devices in either direction. This enables devices to compute an initial score based on the scores computed by mutual peers in the past. For instance,  if device B is not previously known to device A, device A can obtain and compute an initial trust score from its own trusted peers that have also interacted with and therefore have computed trust scores for device B.
    \item \textbf{Common interests}: When both device A and device B have explicitly indicated common interests or are determined using privacy-friendly approaches like private set intersection, then device A determines an input to the trust model for device B based on the common interests. The common interest is learned during the discovery phase both online and offline.
    \item \textbf{Common data or applications}: When both device A and device B have common data or applications installed, device A determines an input weight for computing the trust score for device B based on the common data or application.
    \item \textbf{Proximity data}: The proximity details captured by the wireless sensors is an important factor in determining trust scores. The inputs to the model vary in a non-linear fashion depending on the time, location, frequency and how far apart and how long are devices in proximity to one another during each interaction. Proximity data further includes environment variables in proximity to the devices that can be captured by any array of sensors during each interaction session. 
    \item \textbf{Sensed physical interaction}: Inputs to the trust model are also assigned based on the type and details of any physical interaction between the users captured by their respective on-device sensors.
\end{itemize}

\begin{figure}[htp]
    \centering
    \includegraphics[scale=0.15]{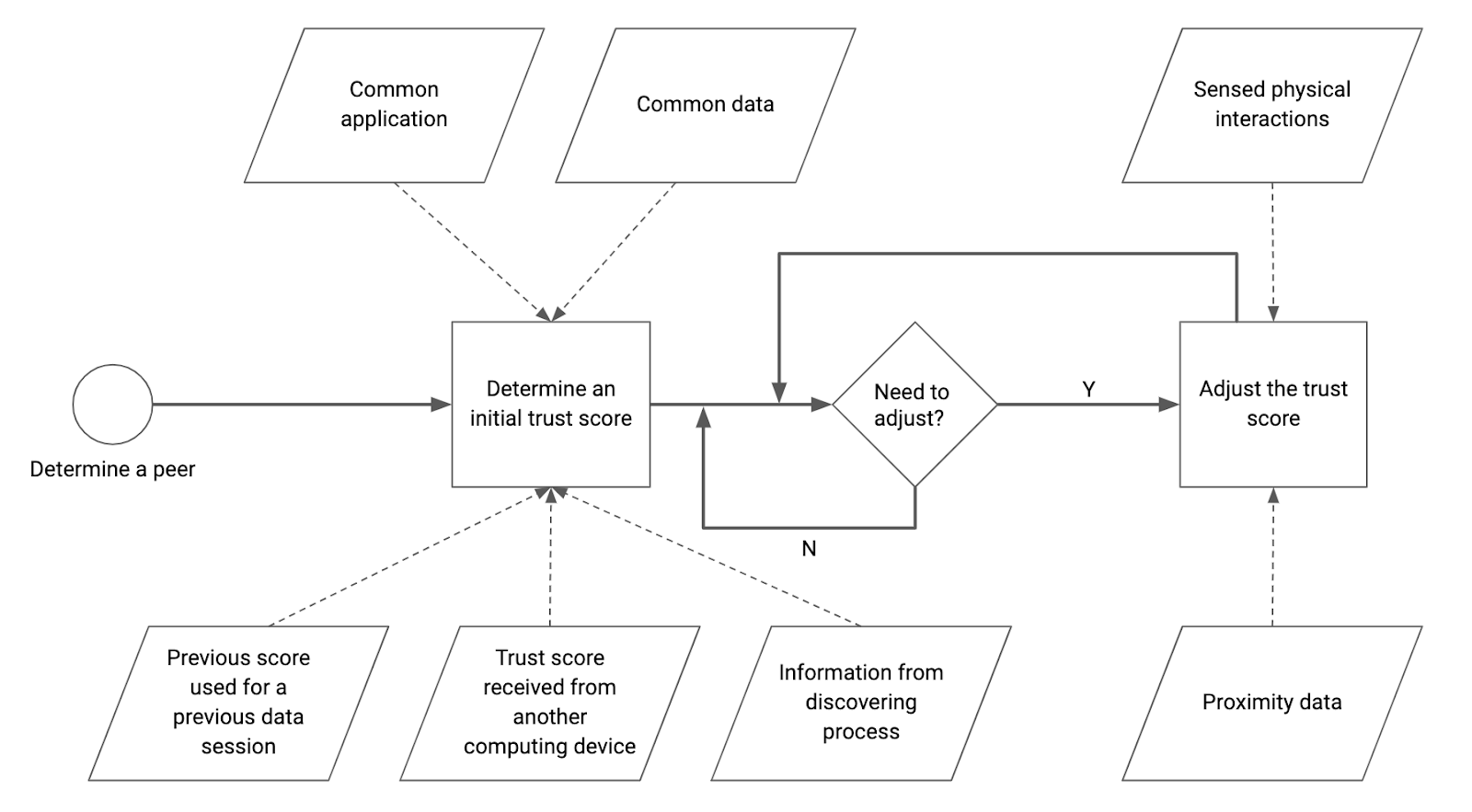}
    \caption{Factors for determining an initial trust score and adjusting based on proximity and interaction data}
    \label{fig:graphic2}
\end{figure}

The trust score and model to determine the scores between any two communicating peers can change over time. For example, at the beginning; device A has a certain trust score for device B based on some or all of the above factors. As device A and device B keep exchanging messages, the trust score for device B can increase and additional factors could be included in determining future scores. On the other hand, when device B has been away for a long period of time, device A may lower the trust score for device B based on the model. 
The trust scores are computed bi-directionally. Because they can be modeled with different evolution and decay functions and include factors that depend on the nature of communication and physical interactions, the scores could be asymmetric.

Further, profiles of trust scores can be created based on context such as nature and process of discovery, interaction type, external events, communication channels, message data, environment variables, etc.

\subsection{Slow Reveal Message Decoding}
The trust score determined on mobile devices is used for secure message passing. Device A issues a security key to device B. Device B uses the security key for decrypting messages based on the trust scores between two devices.
 
When device A sends a security key, it can set a minimum required trust score of the intended receiver for the message. Device A first verifies that the current trust score for device B satisfies the threshold for the message to be transmitted. Device A then sends the message to device B, if the score is above the threshold. 
 
When device B receives the message from device A it determines whether the current trust score for device A satisfies a threshold for the message to be received. In response to the determination, device B decodes the message using the security key received.
 
During the process of message delivery, in addition to setting a threshold for transmission, device A may also set a threshold trust score for reception which would then determine whether the message can be decrypted by the user based on the computed trust score on device B. Because the trust scores change with time based on subsequent interactions, we term the decoding process as “slow reveal” where the decryption of the same message could also be distributed over space and time as appropriate threshold conditions are met between a pair of devices. One way of achieving such a slow reveal would be to have a probabilistic mapping of encrypted input data in different partitions in such a way that each item gets correctly decrypted or decoded as a function of some random variable which could be parametrized by the time profile as well as other needed metrics. The partitioning process allows to reveal only a partial number of bits in a given bit sequence, restricting the receiver from decrypting the complete content.

\begin{figure}[htp]
    \centering
    \includegraphics[scale=0.15]{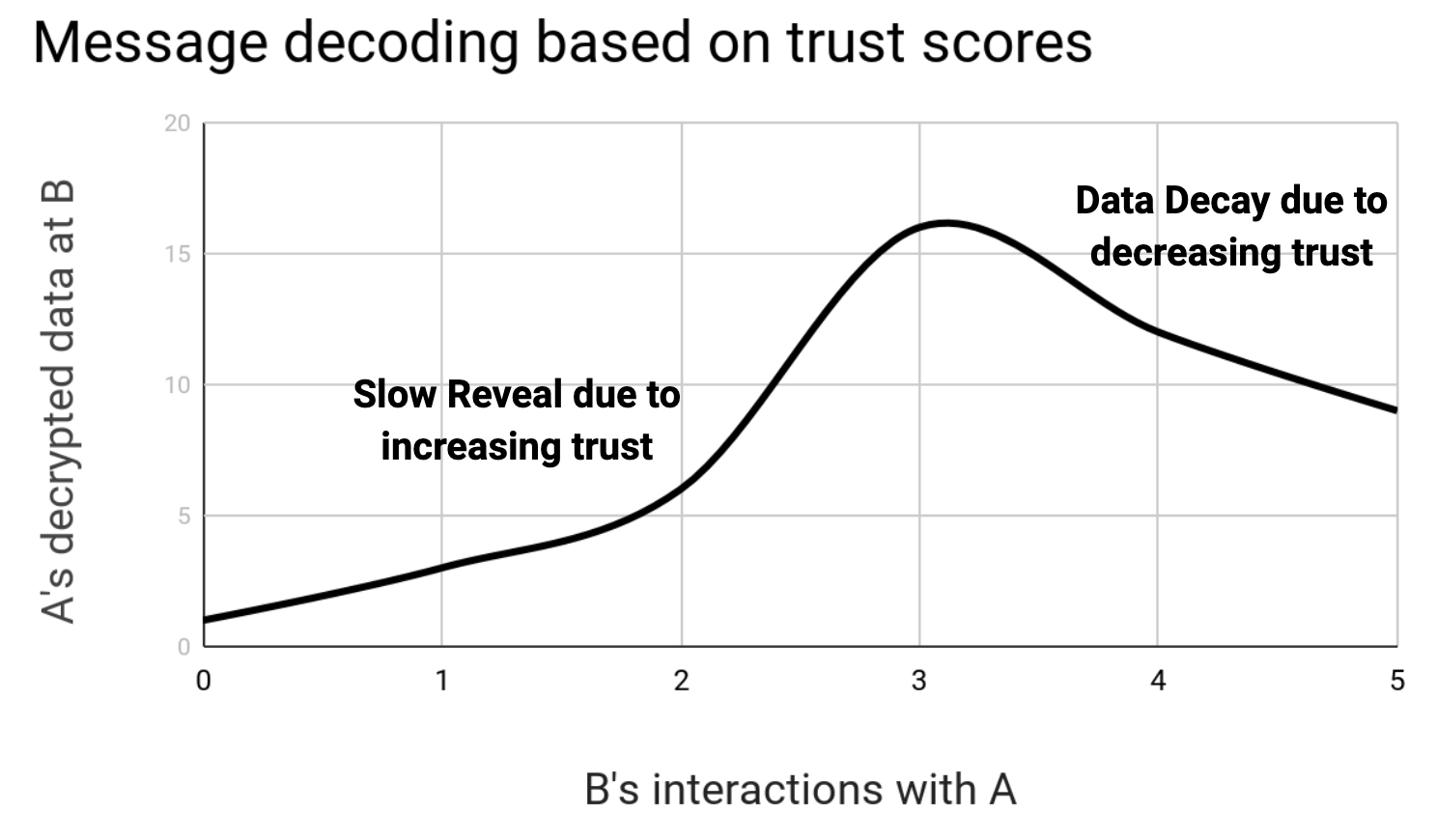}
    \caption{Slow reveal of messages based on trust between A and B}
    \label{fig:graphic3}
\end{figure}

\section{Applications in Epidemics}
The system of establishing trust-based scores can have multi-faceted applications especially in epidemics. Epidemics like COVID-19 can be contained through mapping of peers of infected patients, facilitate safe and informed information exchange of patients and privacy-enabled spread of information among individuals regarding potential susceptibility to the disease. 

\subsection{Contact Tracing - Forward and Backward}
Trust scores are fundamentally calculated as a function of space (proximity) and time which is the same as communicable diseases that result in large outbreaks. \cite{Source27} When the majority of the population has this protocol enabled using Bluetooth based devices combined with GPS and other sensors to provide context like location information, the bi-directional, transitive and asymmetric nature of trust can help in contact tracing, both forward and backward. The scores can be directly used as a proxy in determining transmission rates in disease SEIR models. Epidemiological simulations \cite{Source28} have shown that bi-directional contact tracing can reduce the ambitious adoption requirements for only forward direction based contact tracing tools and hence improve the likelihood of success with relatively lower adoption rates. Once sufficient infected cases have been found, backward tracing can be done using communication based on reception trust score thresholds for specific interaction types that result in disease spreading to ultimately locate patient zero. 

Frequency of interactions and spread of trust scores across peers can further help in locating super spreaders or isolated groups. 

\subsection{Sensitive Data Sharing with Authorized Personnel}
Contact tracing solutions require infected patients to reveal their sensitive personal health and location information to Government authorities and healthcare providers. While numerous privacy-preserving technologies are emerging to ensure that minimal sharing of sensitive information can still provide maximum utility, most of these technologies do not fully protect from data forwarding or leaking to unauthorized people. The most privacy preserving version right now is based from the PACT \cite{Source29} and Google-Apple exposure notification service \cite{Source4} \cite{Source5} \cite{Source6} which does protect the data privacy of individuals. However, there are concerns around other meta-data leakage \cite{Source30}.

Trust scores are computed based on mutual peer scores in a transitive manner. This can enable individuals to verify the authority of government or healthcare personnel through a profile based on message type that has been updated by peer groups. For instance,  when doctors interact with patients, although the interaction might last for a short time, they continue to build their trust scores for exchanging health data as a profile as they see more patients. New patients can set a high reception trust score threshold for health data which is sensitive. In the discovery process they can immediately identify doctors as peers in the network that have high trust scores for health data profiles based on previous interactions with other patients. Only these peers who are doctors can then decode the information on their end preventing any unauthorised access.

\subsection{Privacy-enabled Contextual Information Spreading}
One of the important functions of the government and health authorities during a pandemic is to alert and spread awareness among people who could potentially be at risk based on tested individuals and their contact traces. This must be done carefully to avoid panic by ensuring appropriate messages are issued to the public.  

Proximity-based trust scoring provides both varying degrees of relationships around people and disease susceptibility based on their interactions. This degree of separation from the infected could be effectively used for alerting with different levels of messaging revealing individual or just location information. For instance,  people with highest level of trust scores could be close family or colleagues of the patient who should be informed with specific messaging whereas people with lower trust scores who represent infrequent visitors or passers-by could still potentially be at risk but can be notified at a locality-risk level and encouraged to get tested without revealing any personally identifiable information (PII) of the infected individuals. 

\section{Conclusion}
Proximity-based trust protocol enables utilizing human-like discretionary trust as a factor in digitally connected systems using wireless technologies such as Bluetooth. The trust scores are dynamic, bi-directional, asymmetric, transitive and non-linear between users computed in mobile and interconnected user-held devices. While these properties govern human interaction, behaviours and mobility patterns, in turn pandemic spread of communicable diseases, they can also be used as effective tools in the digital medium to contain such diseases when adopted at scale. We also believe such a protocol can be useful in many other real-world applications where transfer and exchange of goods and information is implicitly or explicitly driven by trust such as private messaging, payments, discovering new friends, and more.

\section*{Acknowledgment}
The authors would like to thank Abhishek Singh (Camera Culture Group, MIT Media Lab) and Rohan Iyer (PathCheck Foundation) for providing inputs and reviewing drafts of this paper. 

\bibliographystyle{IEEEtranN}
\bibliography{References}

\end{document}